\documentclass[aps,prd,preprintnumbers,nofootinbib,twocolumn]{revtex4}
\usepackage{bm}
\usepackage{latexsym}
\usepackage{dcolumn}
\usepackage{amsmath,amsfonts,amssymb}
\usepackage{graphicx,epsfig}
\usepackage{psfrag}
\usepackage{amsthm}

\def\be {\begin{equation}}
\def\ee {\end{equation}}
\def\bea {\begin{eqnarray}}
\def\eea {\end{eqnarray}}
\def\bc {\begin{center}}
\def\ec {\end{center}}
\def\bfg {\begin{figure}}
\def\efg {\end{figure}}
\def\bi {\begin{itemize}}
\def\ei {\end{itemize}}
\def\nn {\nonumber}

\def\la {\label}
\def\le {\left}
\def\ri {\right}
\def\pa {\partial}
\def\fr {\frac}

\def\no {\noindent}

\def\a  {\alpha}
\def\b  {\beta}

\def\D  {\Delta}
\def\e  {\epsilon}

\def\k  {\kappa}
\def\l  {\lambda}

\def\o  {\omega}

\def\th {\theta}

\def\beq{\begin{equation}}
\def\eeq{\end{equation}}
\def\br{\begin{eqnarray}}
\def\er{\end{eqnarray}}
\newcommand{\eel}[1] {\label{#1}\end{equation}}

\newcommand{\bdm}{\begin{displaymath}}
\newcommand{\edm}{\end{displaymath}}

\begin{document}
\title{Effect of the Generalized Uncertainty Principle on Post-Inflation Preheating}

\author{Wissam Chemissany $^1$} \email[email: ]{wissam@itf.fys.kuleuven.be}
\author{Saurya Das $^2$} \email[email: ]{saurya.das@uleth.ca}
\author{Ahmed Farag Ali $^2$} \email[email: ]{ahmed.ali@uleth.ca}
\author{Elias C. Vagenas $^3$}\email[email: ]{evagenas@academyofathens.gr}

\vspace{5ex}
\affiliation{$^1$Instituut voor Theoretische Fysica, Katholieke Universiteit Leuven,\\
Celestijnenlaan 200D, B-3001 Leuven, Belgium\\}

\affiliation{$^2$Theoretical Physics Group, Department of Physics and Astronomy, \\
University of Lethbridge, \\
4401 University Drive,
Lethbridge, Alberta, Canada T1K 3M4 \\}

\affiliation{$^3$~Research Center for Astronomy and Applied Mathematics,\\
Academy of Athens, \\
Soranou Efessiou 4, GR-11527, Athens, Greece\\}

\begin{abstract}
We examine effects of the Generalized Uncertainty Principle, predicted by various theories
of quantum gravity to replace the Heisenberg's uncertainty principle near the Planck scale, on
post inflation preheating in cosmology, and show that it can predict either an increase or a decrease in parametric
resonance and a corresponding change in particle production. Possible implications are considered.
%
%
%
\end{abstract}

\maketitle


\section{Introduction}
\par\noindent
It is generally believed that there was a so-called {\it
preheating} phase near the end of inflation, in which energy was
rapidly transferred from the inflaton to matter fields. This was
followed by {\it reheating}, in which thermalization took place,
and most of the standard model particles in our universe were
produced \cite{Allahverdi:2010xz,Brandenberger:2010dk,Mazumdar:2010sa}.
During preheating, coherent oscillations of the inflaton field
$\varphi(t),$ around the minimum of its potential effectively
contributed to a time-varying frequency term to the equations of
motion of matter fields $\chi(t)$ coupled to it, thereby inducing
instabilities by a well understood process known as {\it
parametric resonance} (PR)
\cite{ll,Allahverdi:1996xc,Kaiser:1995fb,Craig1}. These in turn
resulted in an explosive particle production
\cite{STB,KLS1,KLS2}\footnote{It is also possible PR to result in
a particle production which will not be explosive
\cite{TB,Starobinsky:1981vz}. We thank the anonymous referee for
pointing this out to us.} (see also \cite{Allahverdi:2010xz} for a
recent review). Since the energy density of matter and radiation
is exponentially small near the end of inflation, it has been
argued that this transfer of energy from inflaton to matter fields
must be fast. It was recently shown that (in)homogeneous noise,
such as those arising out of quantum fluctuations, indeed
increased the instability band, and the rate of particle
production \cite{Craig1}. Also it was shown in the past that these
resonance effects were sensitive to non-linearities in the
equations of motion for $\chi$, which in general were expected to
increase the rate of particle production and result in an early
termination of PR \cite{Allahverdi:1996xc}.

In this paper, we completely
analytically study the effect of one such important non-linearity, that predicted from the so-called
{\it Generalized Uncertainty Principle} (GUP). This in turn has been predicted from various approaches
to Quantum Gravity, to replace the familiar Heisenberg's Uncertainty Principle near the Planck scale.
We observe that, depending on the form of the GUP chosen, and initial conditions,
an enhancement of particle production and an early termination of PR can indeed result.
Our paper is organized as follows: in Section II we give a brief review of GUP. In Section III, we review
relevant aspects of PR. Section IV considers effects of general non-linear terms on PR and
Section V the specific ones due to GUP. Section VI generalizes our analysis to an expanding universe. We conclude
in Section VII.

\section{Generalized Uncertainty principle}
\par\noindent
Various approaches to quantum gravity, including String Theory, Loop Quantum Gravity,
as well as Black Hole Physics predict a minimum measurable length of the order of the
Planck length, and a modification of the Heisenberg Uncertainty Principle to a so-called
Generalized Uncertainty Principle, or GUP, near the Planck scale \cite{guppapers,kmm,kempf,brau,sm,cg,subir}.
Such a form of GUP and the corresponding modified commutators between position and momentum coordinates
 was suggested earlier by two of the current authors \cite{dvprl,dvcjp}
\bea
\Delta x_i \Delta p_i &\geq& \fr{\hbar}{2} [ 1 + \beta
\le((\Delta p)^2 + <p>^2 \ri) \nn \\
&+& 2\beta \le( \Delta p_i^2 + <p_i>^2\ri) ]~,~i=1,2,3 \la{uncert1} \\
\le[x_i,p_j \ri] &=& i
\hbar ( \delta_{ij} + \beta \delta_{ij} p^2 + 2\beta p_i p_j ) \la{com1}
\eea
where $p^2 = \sum\limits_{j=1}^{3}p_{j}p_{j}$,
$\beta=\beta_0/(M_{Pl}c)^2=\b_0 \ell_{Pl}^2/2\hbar^2$, $M_{Pl}=$
Planck mass, and $M_{Pl} c^2=$ Planck energy $\approx 1.2 \times
10^{19}~GeV$.
Note that the new ($\b$-dependent) term
is {\it quadratic} in the Planck length (inverse Planck mass).

Recently, some of us proposed another version of GUP consistent
with the above as well as with {\it Doubly Special Relativity}
(DSR) theories \cite{advplb,advplb2}. Some phenomenological
implications of the above GUP were also examined
\cite{advplb,advplb2,Ali:2010yn,Ali:2011fa}
This new version of GUP is of the form
\bea
%
%
 \Delta x \D p \hspace{-1ex}&\geq &\hspace{-1ex}\frac{\hbar}{2}
\le[ 1 - 2 \alpha <p> + 4\alpha^2 <p^2>
\ri] ~ \nn \\
\hspace{-1ex}&\geq& \hspace{-1ex}
\frac{\hbar}{2} \hspace{-1ex}
\le[\hspace{-0.5ex} 1\hspace{-0.5ex}  +\hspace{-0.7ex}  \le(\hspace{-0.7ex}
\frac{\alpha}{\sqrt{\langle p^2 \rangle}} +\hspace{-0.2ex}4\alpha^2 \hspace{-0.9ex} \ri)
\hspace{-0.6ex}  \D p^2 \hspace{-0.6ex}
+\hspace{-0.6ex}  4\alpha^2 \langle p \rangle^2 \hspace{-0.6ex}
- \hspace{-0.6ex}  2\alpha \sqrt{\langle p^2 \rangle}\hspace{-0.2ex}
\ri]\hspace{2ex} \label{dxdp1} \\
\le[x_i, p_j \ri]\hspace{-1ex} &=&\hspace{-1ex} i \hbar\hspace{-0.5ex} \left[  \delta_{ij}\hspace{-0.5ex}
- \hspace{-0.5ex} \alpha \hspace{-0.5ex}  \le( p \delta_{ij} +
\frac{p_i p_j}{p} \ri)
+ \alpha^2 \hspace{-0.5ex}
\le( p^2 \delta_{ij}  + 3 p_{i} p_{j} \ri) \hspace{-0.5ex} \ri]
\label{comm01}
\eea
where
$\alpha = {\alpha_0}/{M_{Pl}c} = {\alpha_0 \ell_{Pl}}/{\hbar}.$
%
%
Note that in this case, the new ($\a$-dependent) term
is {\it linear} in the Planck length (inverse Planck mass).
\par\noindent
It is also noteworthy that although Eqs. (\ref{dxdp1}) and (\ref{comm01}) are not
Lorentz covariant, they are
DSR covariant \cite{cg}. We expect the results of our paper
to have similar covariance as well. In addition, since DSR transformations
preserve not only the speed of light, but also the Planck momentum
and the Planck length, it is not surprising that Eqs.
(\ref{dxdp1}) and (\ref{comm01}) imply the following minimum
measurable length {\it and} maximum measurable momentum
\bea
\D x &\geq& (\D x)_{min}  \approx \alpha_0\ell_{Pl} \la{dxmin} \\
\D p &\leq& (\D p)_{max} \approx \frac{M_{Pl}c}{\alpha_0}~. \la{dpmax}
\eea
\par\noindent
It is evident that the GUP induced terms in both
(\ref{com1}) and (\ref{comm01}) become important near the Planck scale.
Also they both ensure, via the Jacobi identity, that
$[x_i,x_j]=0=[p_i,p_j]$.
\par\noindent
Next, defining
\bea x_i &=& x_{0i}~,~~
p_i = p_{0i} \le( 1 + {\beta} p_0^2 \ri) \la{mom1a} \\
\mbox{and~}~~x_i &=& x_{0i}~,~~
p_i = p_{0i} \le( 1 - \alpha p_0 + 2\alpha^2 p_0^2 \ri)~
\la{mom1b}
\eea
where $p_0^2 = \sum\limits_{j=1}^{3}p_{0j}p_{0j}$ and $x_{0i}, p_{0j}$
satisfying the canonical commutation relations
%
$ [x_{0i}, p_{0j}] = i \hbar~\delta_{ij}, $
%
it is easy to show that equations (\ref{com1}) and (\ref{comm01}) are satisfied
to order $\beta$ and $\a^2$, respectively.
%
Here,
$p_{0i}$ can be interpreted as the momentum at low energies
(having the standard representation in position space, i.e. $p_{0i} = -i
\hbar d/d{x_i}$), and $p_{i}$ as that at higher energies.
\par\noindent
Finally, employing Eqs. (\ref{mom1a}) and (\ref{mom1b}), it is easy to see that
a non-relativistic Hamiltonian of the form
\bea H &=& \fr{p^2}{2m} + V(\vec r)
~~~[\vec r =(x_1,x_3,x_3)]
\la{ham1}
\eea
translates to
\bea H&=&
H_0 + H_1 + {\cal O}( \beta^2) ~~\mbox{or}~~  {\cal O}(\a^2) ~, \\
%
%
\mbox{where}~~H_0 &=& \fr{p_0^2}{2m} + V(\vec r) \nn \\
\mbox{and}~~ H_1 &=& \fr{\beta}{m} p_0^4 ~~\mbox{or}~~
H_1 = -\frac{\a}{m}p_0^3~
\la{H0}
\eea
for the different versions of GUP defined by equations (\ref{com1}) and (\ref{comm01}), respectively.
Thus, we see that {\it any} system with a
well defined quantum (or even classical) Hamiltonian $H_0$, is perturbed
by $H_1$, defined above, near the Planck scale.
It may be mentioned that it is often assumed that the dimensionless GUP parameters,
$\alpha_0$ and $\beta_0$ are of order unity.
Also, the momenta $p$ are expected to be much less than $M_{pl}c/\a_0$ or $M_{pl}c/\sqrt{\b_0}$, because of the immensity of the Planck scale.
Therefore the higher order terms which have been left out are sub-leading and unimportant.
As for the leading order terms that have been retained, the resulting effects although small, may be perceptible, which as shall
see in later sections, may be further enhanced by various kinematical factors
%
\footnote{
Recently, three of the current authors,
without assuming the strengths of $\a_0, \b_0$ {\it a priori}, derived
bounds on the linear GUP parameter \cite{Ali:2011fa} from
experimental accuracies of various quantum systems such as the Landau levels, the Simple Harmonic Oscillator, the Lamb Shift, and the tunneling current in a Scanning Tunneling Microscope (STM) etc.
From the first three quantum phenomena,
the upper bounds on $\alpha_0$ were found to be $10^{23}$, $10^{17}$ and $10^{10}$, respectively.
The first bound leads to a length scale bigger than the electroweak length scale,
and so although not inconsistent with it, does not lead to any new verifiable scale.
The second and third bounds could signal a new and intermediate length scale between the electroweak and the Planck scale.
It was also found that even for $\alpha_0 \approx 1$, one might be able to
detect quantum gravitational corrections in an STM.
Additionally, the bounds on the quadratic GUP parameter \cite{dvprl,dvcjp},
for the Lamb Shift, Landau Levels and STM
were found $10^{18}$, $10^{25}$ and $10^{10}$, respectively.
In other words, the bounds derived for the linear GUP parameter are more stringent than those derived for the quadratic GUP parameter.
It is possible that suitable astrophysical or cosmological observations may also put some useful bounds on the
above parameters.
}.
%
%
%
\section{Particle production: Parametric Resonance}
%
%
\par\noindent
We review and closely follow the analysis of \cite{ll,Craig1}, and
consider an oscillating scalar field
\begin{equation}\varphi(t)=h \cos(\th t)\end{equation}
coupled to another scalar field $\chi$ representing the matter, via an interaction
\begin{equation}
\mathcal{L} \propto \frac{1}{2} ~\varphi \chi^{2}~.
\end{equation}
Then the evolution equation for $\chi$ is nothing but the Mathieu equation,
of the form
\be
\ddot {\chi} + \o_0^2 \Big(1 + h \cos\le[(2\o_0 + \e) t\Big] \ri)\chi =0~
\la{eom1}
\ee
where the argument of the cosine, $\theta \equiv 2\o_0 + \e$,
is so chosen to produce the strongest parametric
resonance (PR) via the $h$-term. We shall assume $h,\e/\o_0<<1$,
and retain terms only to leading order in $h,\e$.
Assuming a solution of the form
\be
\chi(t) = a_0 (t) \cos\left(\frac{\th}{2} t\right) + b_0 (t)
\sin\left(\frac{\th}{2} t\right)
\la{soln1}
\ee
with $a_0 \sim e^{s_0t},~b_0 \sim e^{s_0 t}$, substituting in Eq. (\ref{eom1}),
using $\dot a_0 \sim \e a_0, \dot b_0 \sim \e b_0$ (thereby ignoring
$\ddot a_0 \sim \e^2 a_0$ and $\ddot b_0 \sim \e^2 b_0$ terms),
identities such as
$\cos A \cos B = \frac{1}{2} [ \cos (A+B) + \cos (A-B) ]$
and
$\cos A \sin B = \frac{1}{2} [ \sin (A+B) - \sin (A-B) ]$,
 and ignoring weaker resonance
terms of the form $\cos(n\frac{\th}{2} t)$ and
$\sin(n\frac{\th}{2} t)$ ($n \in \mathbb{N}>1$), we get
\bea
a_0 s_0  + \frac{b_0}{2} \le( \frac{h\o_0}{2} + \e \ri) &=& 0 \la{a1} \\
b_0 s_0  + \frac{a_0}{2} \le( \frac{h\o_0}{2} - \e \ri) &=& 0~. \la{b1}
\eea
Solving the above two equations, we obtain
\bea
&&  \frac{b_0}{a_0} = \sqrt{\frac{\frac{h\o_0}{2} - \e}{ \frac{h\o_0}{2} + \e  }} \equiv R \label{RE}\\
&& s_0 = \frac{1}{2} \sqrt{(\frac{1}{2} h\o_0)^2 - \e^2 } ~.
\label{unperturbed-s0}
\eea
Note that $0 \leq R < \infty$, with $R=1$ corresponding to $a_0=b_0$ and
$R=0~(\infty)$ corresponding to $b_0=0~(a_0=0)$.
Thus, when $s_0 \in \mathbb{R}$, the solution given in Eq. (\ref{soln1}) grows exponentially
in the so-called {\it Instability Region} of $h\o_0$
given by the parameter range
\be
-\frac{1}{2} h \o_0 < \e  < \frac{1}{2} h \o_0~.
\label{width1}
\ee
This is the phenomenon of Parametric Resonance \cite{ll}.
%
%
%
%
\section{Parametric Resonance with Non-Linear Terms}
%
%
\par\noindent
We introduce a generic non-linearity in the RHS of Eq. (\ref{eom1}) of the following form
\be
\ddot {\chi} + \o_0^2 \Big( 1 + h \cos\le[(2\o_0 + \e) t\ri]\Big) \chi = \l f(\chi, \dot \chi)~
\la{eom2}
\ee
where $f(\chi, \dot \chi)$ is an arbitrary non-linear function.
As we shall see later, $\lambda$ is suppressed by powers of Planck mass, and its effects may only
show up at very high energies and very small length scales.
We will thus treat this term perturbatively.
Once again, we assume  Eq. (\ref{eom2}) has a solution of the form
\bea
\chi(t) &=& a (t) \cos\left(\frac{\th}{2} t\right) + b (t)
\sin\left(\frac{\th}{2} t\right),
\la{soln2}\\
&&\mbox{where}~~~a\sim  e^{st}~, b \sim  e^{st}
\eea
with $a = a_0 + \l a_1, b = b_0 + \l b_1, s = s_0 + \l s_1$~.
\par\noindent
Thus we can substitute the `unperturbed' solution given by Eq. (\ref{soln1}) in the RHS of Eq. (\ref{eom2}),
simplify again using trigonometrical identities, and retain only the leading order resonance terms to
write
\bea
\l f(\chi,\dot \chi) = \l \o_0 \sin\left(\frac{\th}{2} t\right) f_1 (a_0,b_0, s_0) \nn \\
+ \l \o_0 \cos\left(\frac{\th}{2} t\right) f_2 (a_0,b_0,s_0)~. \la{f1f2} \eea
This modifies Eqs.(\ref{a1}-\ref{b1}) to
\bea
a s  + \frac{b}{2} \le( \frac{h\o_0 }{2} + \e \ri) &=& - \frac{\l}{2} f_1 \la{a2} \\
b s  + \frac{a}{2} \le( \frac{h\o_0}{2} - \e \ri) &=&  \frac{\l}{2} f_2~. \la{b2}
\eea
Now we make a further assumption about the smallness of $\l$,
namely that $\l \sim \e^{1+p},~0\leq~p<1$ and retain terms up
to ${\cal O}(\e^{2})$ to obtain
%
%
%
\bea
s^2 = s_0^2 + \frac{\l}{4}
\le[
 \frac{f_1}{b_0} \le(\frac{h\o_0}{2} - \e \ri)
 - \frac{f_2}{a_0} \le(\frac{h\o_0}{2} + \e \ri)
\ri]~.
\eea
Comparing with the leading order expression $s^2 = s_0^2 + 2 \l s_0 s_1$ we get
\bea
2 s_0 s_1 &=& \frac{1}{4}
\le[
\frac{f_1}{b_0} \le(\frac{h\o_0}{2} - \e \ri)
- \frac{f_2}{a_0} \le(\frac{h\o_0}{2} + \e \ri)
\ri] \\
&=&
\frac{1}{4a_0}
\le(\frac{h\o_0}{2} + \e \ri)
\le[
R f_1 - f_2
\ri]~.
\la{s0}
\eea
Thus to determine whether PR is enhanced ($s_1>0$) or diminished ($s_1<0$), one would
need to find expressions for $f_1,f_2$ for specific models of non-linearity.
%
%
%
\section{Parametric Resonance with GUP}
%
%
%
%
\par\noindent
We start with a GUP-modified Hamiltonian in one dimension of the form
\bea H = \frac{p^2}{2m} + V(\chi) = \frac{p_0^2}{2m} + V(\chi) +
\frac{\k \alpha^{n-2}}{m}p_0^n~, \label{Ham}\eea
where $\k=\pm 1$. This incorporates the two
versions of GUP presented in Section II, with $n=4$ and $n=3$, respectively,
(cf. Eq. (\ref{H0})).
The first of the equations of Hamilton, and its inverse, both to ${\cal O}(\alpha^{n-2})$ are
given by
\bea
\dot \chi &=& \frac{\pa H}{\pa p_0} = \frac{p_0}{m} + \frac{\k n \alpha^{n-2}}{m} p_0^{n-1} \\
p_0 &=& m \dot \chi - n \k \alpha^{n-2} (m \dot \chi)^{n-1}~. \eea
Then the second Hamilton equation $\dot p_0 = -\pa H/\pa \chi$ gives
\bea
m \ddot \chi \le[ 1 - \k n(n-1) \alpha^{n-2} m^{n-2} \dot \chi^{n-2} \ri] &=& - \frac{\pa V}{\pa \chi} \\
\mbox{or,}~~
m \ddot \chi + \frac{\pa V}{\pa \chi} + \k_1 \frac{\pa V}{\pa \chi} \dot \chi^{n-2} &=& 0
\la{eom4}
\eea
where $\k_1 = \k n(n-1) \alpha^{n-2} m^{n-2}$. Thus for the time
dependent harmonic oscillator, we get from Eq. (\ref{eom4}) above
\bea \ddot \chi + \o_0^2 \Big( 1+h\cos\le(\th t\right)\Big)\chi = -\k_1 \o_0^2 \chi \dot
\chi^{n-2}~
\label{eqn36}
\eea
where we have ignored terms ${\cal O}(\k_1 h)$.
Thus , comparing Eqs. (\ref{eom2}) and (\ref{eqn36}), we obtain $\l f = -\k_1
\o_0^2 \chi \dot \chi^{n-2}$. Further, we will make the identification
$\l=\alpha^{n-2}$. Next we write the solution given in Eq. (\ref{soln2}) as
\bea
\chi &=& c e^{i\frac{\th}{2} t} + c^\star e^{-i\frac{\th}{2} t} \\
\dot \chi &=& A e^{i\frac{\th}{2} t} + A^\star e^{-i\frac{\th}{2} t} \\
c &=& \frac{1}{2} \le( a + \frac{b}{i} \ri)~,~~ A = \frac{1}{2}
\le[ (\dot a + \frac{\th}{2} b) + \frac{\dot b - \frac{\th}{2}
a}{i} \ri] \eea
from which the quantity $\chi \dot\chi^{n-2}~$ that appears in the RHS of Eq.  (\ref{eqn36}) reads
\bea && \chi \dot \chi^{n-2} =
\le( c e^{i\frac{\th}{2} t} + c^\star e^{-i\frac{\th}{2} t} \ri) \le( A e^{i\frac{\th}{2} t} + A^\star e^{-i\frac{\th}{2} t} \ri)^{n-2} \\
&&= \sum_{j=0}^{n-2} {n-2 \choose j} A^j A^{\star (n-2-j)}
\Big( c e^{i(2j-n+3)\frac{\th}{2} t}+\\
 &&  c^\star
e^{i(2j-n+1)\frac{\th}{2} t}
 \Big)~.
\eea
Setting $2j-n+3 = \pm1$ and $2j-n+1 = \pm 1$ to extract the
$e^{\pm i\frac{\th}{2} t}$, i.e. the dominant resonance terms, we
see $j=(n-2)/2,(n-4)/2$ and $j=n/2,(n-2)/2$, respectively.
It is evident that $n$ must be even and for the rest of the paper, we assume that this
is the case\footnote{Higher order (weaker) resonance terms can of
course arise for $n$ odd.}.
Collecting these terms, simplifying, using $\dot a = sa,~\dot b = sb$, and replacing
$\{a,b,s\}$ by $\{a_0,b_0,s_0\}$ by noting that the above term is multiplied by
the small parameter $\l$
in the equation of motion, and finally comparing with Eq. (\ref{f1f2}) we get
\bea && f_1 =\label{fone} \ell a_0^3 \le[ \frac{n
}{n-2}(4s_0^2+\theta^{2})R+R(4s_0^2-\th^2)-4 s_0 \th
\ri]~~~~ \\
&& \label{ftwo}f_2 =\ell a_0^3\le[ \frac{n
}{n-2}(4s_0^2+\theta^{2})+(4s_0^2-\theta^{2}) +4 Rs_0\th
\ri] \\
&&\mbox{where}~~~\ell = \frac{\k_2  (1+R^2)}{16}\\
&&\mbox{and}~~~\k_2 = -\k n (n-1) m^{n-2}\o_0 {{n-2}\choose {n/2}} |A|^{n-4}\nonumber\\
\eea
which using Eq. (\ref{s0}), implies in terms of the ratio $R$,
\begin{eqnarray}  \label{sszero}2s_0 s_1 &=& \frac{\ell a_0^2}{4}\Big(\frac{h\o_0}{2}+\e\Big)
 \Big[(R^2 - 1)\Big(\frac{n}{n-2}(4
s_0^2+\th^2)\nonumber\\&&+(4s_0^2-\th^2)\Big)-8 R s_0\th\Big]~.
\end{eqnarray}
\par\noindent
We use the last equation to summarize the sign of $s_1$ (Note that $s_0$ is positive)
\begin{center}
  \begin{tabular}{ | c || c | c| }
    \hline
     & sign of $s_1$ & sign of $s_1$  \\
     & ($\kappa>0 $) & ($ \kappa < 0 $) \\
     \hline \hline
    $R=1$  &  $+$ & $-$ \\
\hline
    $R=0$ &
$ + $ & $-$ \\
\hline
$R=\infty$ & $+$ & $-$  \\
\hline
$R \neq 1$ & $\pm$ & $\pm $
\\
\hline
  \end{tabular}
\end{center}
%
%
%
Thus, we see that the initial conditions on the matter field (via $R$)
and the GUP one is considering (via the sign of $\kappa$)
determine whether there is an increase in the exponent of the matter field or not. The
various auxiliary variables that were introduced (such as $A$, $C$, $\ell$ etc)
do not play any role in it.

%
%
\subsection{Instability Region}
%
%
%
\par\noindent
Now setting $s=0$ in Eqs. (\ref{a2}) and (\ref{b2}), we see that the PR occurs when the modified
instability region is given by
\bea
-\frac{1}{2} h \o_0 - \frac{\l f_1}{b_0} < \e  < \frac{1}{2} h \o_0 - \frac{\l f_2}{a_0}
\eea
whose width is thus
\begin{eqnarray}
 \Delta \e&=&h \o_0 - \frac{\l}{a_0 R} \le(R f_2-f_1 \ri)\nonumber\\
&=& h\o_0 - \l \Big[\frac{4(1+R^2)}{R}s_0\th\ell\Big]a_0^{2}\nonumber\\
&=& h\o_0 - 2\l h \o_0\th\ell a_0^{2}\nonumber\\
&=& h\o_0 \Big[1- 2\l \th\ell a_0^{2}\Big]~.
\label{gup}
\end{eqnarray}
\par\noindent
It follows that parametric resonance is maintained in the presence of GUP.
However the region of instability increases when $\kappa>0$ and vice-versa,
i.e. it once again depends on the GUP under consideration,
as well as the amplitude and oscillation frequency of the matter field, which may contribute to further enhancement of this region
(but not on $R$).
However, since such an increase or decrease would be proportional
to inverse powers of the Planck mass, it may turn out to be too
small to have an observable effect at present, although with more
accurate experiments and improved observations, it may be
detectable in the future.
In the next section, we re-do the analysis for an expanding universe.

%
%
%
\section{Parametric Resonance in an expanding universe with nonlinear terms}
%
%
%
\par\noindent
We adopt the procedure outlined in \cite{Kaiser:1995fb,ll}. We first
write the equation for the matter field in the presence of the inflaton field $\varphi$
and in an expanding background \cite{Allahverdi:2010xz}, together with the non-linear terms described by
Eq. (\ref{Ham}) as
\begin{equation}
\label{Phu}
\ddot{\chi}+3H \dot{\chi}+\omega_{0}^{2}\left[1+h
 \cos(2\omega_{0}+\epsilon)t\right]\chi=\lambda f(\chi,\dot{\chi}),
\end{equation}
where
$H=\dot{a}/a$ is the Hubble parameter and $a(t)$ is the scale factor for FRW spacetime.
%
The re-definition
 \begin{equation}Y=a^{3/2} \chi\end{equation}
 implying
 \begin{eqnarray}\label{transformed}\dot{\chi}&=&a^{-3/2}(\dot{Y}-\frac{3}{2}H
 Y)\\ \label{transformed1}
 \ddot{\chi}&=&a^{-3/2}(\ddot{Y}+\frac{9}{4}H^{2}Y-3H\dot{Y}-\frac{3}{2}\dot{H}Y)
 \end{eqnarray}
%
substituted into (\ref{Phu}) yields the following equation for the function $Y$
 \begin{equation}\label{Yeq}\ddot{Y}+\omega^{2} Y=\lambda P[Y,\dot{Y},a]
\end{equation}
 where
 \begin{eqnarray}\omega^{2}(t)&=&\omega_{0}^{2}(1+h
 \cos\theta t)-\frac{9}{4}H^{2}-\frac{3}{2}\dot{H}\nonumber\\
 &&\mbox{with}~~\theta=2\omega_0 +\epsilon~,\\
 \mbox{and}~~P&=& a^{3/2}f[a^{-3/2} Y,a^{-3/2}(\dot{Y}-\frac{3}{2}H Y)]~.
 \end{eqnarray}
Next we consider the following cases.
%
%
%
\subsection{Case when $H\neq0$ and $\lambda=0$}
%
%
%
To gain some insight we  first revisit the case where there is
no nonlinearity. We assume a solution of the form
 \begin{equation}
 \label{sol}
 Y(t)=c_{0}
 \zeta_{0}(a)\cos\left(\frac{\theta}{2}t\right)+d_{0}
 \zeta_{0}(a)\sin\left(\frac{\theta}{2}t\right)
 \end{equation}
in which the effects of expansion are included in the
scaling function $\zeta_{0}(a)$.
Substituting into Eq. (\ref{Yeq}) with its RHS set to zero, one obtains
%
\begin{equation}\label{coefs}(A_{1}+A_{2})
\cos\left(\frac{\theta}{2}t\right)+
(B_{1}+B_{2})\sin\left(\frac{\theta}{2}t\right)+\mathcal{O}(\theta^{2},h^{2})=0\end{equation}
with
\begin{eqnarray}
\label{AABB}
A_{1}&=&\!(\dot{d}_{0}\theta+\omega_{0}^{2}c_{0}+\frac{\omega_{0}^{2}h}{2}c_{0}-\frac{c_{0}\theta^{2}}{4})\zeta_{0}\\
B_{1}&=&\!(-\dot{c}_{0}\theta+\omega_{0}^{2}d_{0}-\frac{\omega_{0}^{2}h}{2}d_{0}-\frac{d_{0}\theta^{2}}{4})\zeta_{0}\\
A_{2}&=&\!\!2\dot{c}_{0}\dot{\zeta}_{0}+c_{0}\ddot{\zeta}_{0}+d_{0}\dot{\zeta}_{0}\theta-\frac{9}{4}H^{2}c_{0}\zeta_{0}-\frac{3}{2}\dot{H}c_{0}\zeta_{0}\\
\label{B2}
B_{2}&=&\!\!2\dot{d}_{0}\dot{\zeta}_{0}+d_{0}\ddot{\zeta}_{0}-c_{0}\dot{\zeta}_{0}\theta-\frac{9}{4}H^{2}d_{0}\zeta_{0}-\frac{3}{2}\dot{H}d_{0}\zeta_{0}~.
\end{eqnarray}
If  Eq. (\ref{coefs}) is to be justified, the coefficient of the sine and cosine must vanish.
In addition, in order to ensure resonant behavior, we further set
$A_{1},\,A_{2},\,B_{1},$ and $B_{2}$ separately equal to zero.
Thus, up to to order $\mathcal{O}(\epsilon,h),$ the coefficients
$A_{1}$ and $B_{1}$  reduce to the Eqs. (\ref{a1}) and (\ref{b1}), i.e.,
\begin{eqnarray}
A_{1}&=& 2s_{0}d_{0}+c_{0}\Big(\frac{h\omega_{0}}{2}-\epsilon\Big)=0\\
B_{1}&=&-2s_{0}c_{0}-d_{0}\Big(\frac{h\omega_{0}}{2}+\epsilon\Big)=0
\end{eqnarray}
where $s_0$ is the characteristic (`unperturbed') exponent defined in Eq. (\ref{unperturbed-s0}).
On the other hand, $A_{2}=0$ and $B_{2}=0$ yield
\begin{eqnarray}
2\dot{c}_{0}\Big(\frac{\dot{\zeta}_{0}}{\zeta_{0}}\Big)+c_{0}\Big(\frac{\ddot{\zeta}_{0}}{\zeta_{0}}\Big)+d_{0}\Big(\frac{\dot{\zeta}_{0}}{\zeta_{0}}\Big)\theta
-\Big(\frac{9}{4}H^{2}+\frac{3}{2}\dot{H}\Big)c_{0}&=&0\nonumber\\
2\dot{d}_{0}\Big(\frac{\dot{\zeta}_{0}}{\zeta_{0}}\Big)+d_{0}\Big(\frac{\ddot{\zeta}_{0}}{\zeta_{0}}\Big)-c_{0}\Big(\frac{\dot{\zeta}_{0}}{\zeta_{0}}\Big)\theta
-\Big(\frac{9}{4}H^{2}+\frac{3}{2}\dot{H}\Big)d_{0}&=&0\nonumber
\end{eqnarray}
which can be rewritten as
\begin{eqnarray}
\frac{1}{\zeta_{0} c_{0}}\frac{d}{dt}(c_{0}^{2}\dot{\zeta}_{0})+d_{0}\Big(\frac{\dot{\zeta}_{0}}{\zeta_{0}}\Big)\theta
-\Big(\frac{9}{4}H^{2}+\frac{3}{2}\dot{H}\Big)c_{0}&=&0\\
\frac{1}{\zeta_{0} d_{0}}\frac{d}{dt}(d_{0}^{2}\dot{\zeta}_{0})-c_{0}\Big(\frac{\dot{\zeta}_{0}}{\zeta_{0}}\Big)\theta
-\Big(\frac{9}{4}H^{2}+\frac{3}{2}\dot{H}\Big)d_{0}&=&0~.
\end{eqnarray}
The two equations above may be combined to give (setting $c_0=C_0e^{s_{0}t}$ and $d_0=D_0e^{s_{0}t}$
in which $C_0$ and $D_0$ are constants)
\begin{equation}
\label{feq}
\frac{d}{dt}(e^{2s_{0}t}\dot{\zeta}_{0})-\left(\frac{9}{4}H^{2}+\frac{3}{2}\dot{H}\right)\zeta_{0} e^{2s_{0}t}=0~.
\end{equation}
This equation may be solved following the procedure given in \cite{Kaiser:1995fb}.
For a cosmic-time scale factor of the form $a(t)\propto t^{q}$,
we have\footnote{It is noteworthy that $\alpha(q)\geq 0$ for
$q\geq 1,$ and that for de Sitter expansion, $\alpha(\infty)=\infty.$}
\begin{equation}
\dot{H}+H^{2}=q(q-1)t^{-2}=\alpha(q)t^{-2},\qquad
H=q t^{-1},
\end{equation}
then Eq. (\ref{feq}) becomes
\begin{equation}
\ddot{\zeta}_{0}+2s_{0}\dot{\zeta}_{0}-\beta(q)\,t^{-2}\zeta_{0} =0~
\end{equation}
where we have introduced the coefficient
$\beta(q)=\frac{3}{4}q^{2}+\frac{3}{2}\alpha(q).$ If we use the
fact that $s_{0}$ is of order $\epsilon$ and $h,$ then we are allowed to approximate  the above equation to
\begin{equation}
\ddot{\zeta}_{0}-\beta(q) t^{-2} \zeta_{0}\simeq 0~.
\end{equation}
The solution to this equation takes the form
\begin{eqnarray}
\zeta_{0}(t)&=&C_{1}t^{\frac{1}{2}(1+\sqrt{1+4\beta})}+
C_{2}t^{\frac{1}{2}(1-\sqrt{1+4\beta})}\nonumber\\
&=&C_{1}t^{\frac{3q}{2}}+C_{2}t^{\frac{2-3q}{2}}\label{solf}~.
\end{eqnarray}
With this solution for $\zeta_{0}(t),$ the resonant solution reads
\begin{equation}
Y^{\pm}(t)=C_{0}\zeta_{0}(t) e^{\pm s_{0}t}\left[\cos
\left(\frac{\theta}{2}t\right)
\mp R \sin\left(\frac{\theta}{2}t\right)\right]
\end{equation}
where $R$ stands for $d_{0}/c_{0}$.
\par\noindent
Following the normalization chosen in \cite{TB}, $C_{0}$ has
been determined to be $C_{0}=\sqrt{\frac{1}{R\theta}}.$ By comparing the
solution given by Eq. (\ref{solf}) to the solution associated with the
nonresonant case as given in \cite{Kaiser:1995fb}, we infer that the second exponent in
Eq. (\ref{solf}) must be neglected. Thus, the appropriate exponent for
the resonant case is the first one. The requirement that
$\zeta_{0}[a(t)]=1$ when $a(t)=1$ implies the normalization $\zeta_{0}=a(t)^{3/2}.$
Therefore, the full resonant solution in an expanding
background behaves as
\begin{equation}
Y^{\pm}(t)=\sqrt{\frac{1}{R\theta}}e^{ s_{\pm}t}\left[\cos
\left(\frac{\theta}{2}t\right) \mp R \sin\left(\frac{\theta}{2}t\right)\right]
\end{equation}
with the characteristic exponent $s_{\pm}(t)$ defined by
\begin{equation}
s_{\pm}(t)=\pm s_{0}+\frac{3q}{2 t}\ln t ~.
\end{equation}
Clearly, the stability band width will decrease after taking the
expanding background into account.  In particular, following Eq. (\ref{width1}) the
bounds of the Instability Region in a flat spacetime are
\begin{equation}
\epsilon_{min}=-\sqrt{\left(\frac{h\omega_0}{2}\right)^2}~~\mbox{and}~~
\epsilon_{max}=+\sqrt{\left(\frac{h\omega_0}{2}\right)^2}
\end{equation}
while when the background is an expanding one as described above the bounds are modified as follows
\cite{ll,Kaiser:1995fb,STB}
\begin{eqnarray}
\epsilon_{min}&=&-\sqrt{\left(\frac{h\omega_0}{2}\right)^2 - 4 \left(\frac{3q}{2 t}\ln t\right)^2}\label{backg1}\\
\epsilon_{max}&=&+\sqrt{\left(\frac{h\omega_0}{2}\right)^2- 4  \left(\frac{3q}{2 t}\ln t\right)^2}\label{backg2}~.
\end{eqnarray}
Now we are in a position to generalize the above calculation to
the nonlinear model governed by Eq. (\ref{Yeq}), and find
the analogue of Eq. (\ref{feq}).
%
%
%
\subsection{Case when $H\neq0$ and $\lambda \neq0$}
%
%
%
\par\noindent
We reconsider equation (\ref{Yeq})  with the nonlinear terms included
 \begin{equation}
 \label{Yeq1}
 \ddot{Y}+\omega^{2} Y=\lambda P[Y,\dot{Y},a]
\end{equation}
and thus, the solution will now be of the form
\begin{equation}
\label{sol1}
Y(t)=c(t)\zeta(a)\cos\left(\frac{\theta}{2}t\right)+
d(t)\zeta(a)\sin\left(\frac{\theta}{2}t\right)
 \end{equation}
\par\noindent
with $c=c_{0}+\lambda c_{1},~d=d_{0}+\lambda d_{1},~s=s_{0}+\lambda s_{1},\zeta=\zeta_{0}+\lambda \zeta_{1}$~.
\par\noindent
Next we substitute the unperturbed solution given in Eq. (\ref{sol})
in the right-hand side of Eq. (\ref{Yeq1}), and as in the previous sections
 perform the trigonometrical approximations and retain
term of order $\mathcal{O}(h).$  We thus write
\begin{eqnarray}
\label{PP}
\lambda P[Y,\dot{Y},a]&=&\lambda\omega_{0}\sin\left(\frac{\theta}{2}t\right) ~P_{1}[c_{0},d_{0},s_{0},\zeta_{0},a]
+\nonumber\\&&\lambda\omega_{0}
\cos\left(\frac{\theta}{2}t\right)~P_{2}[c_{0},d_{0},s_{0},\zeta_{0},a]~.
\end{eqnarray}
Then we obtain up to order $\mathcal{O}(h,\lambda)$
\begin{eqnarray}
(B-\lambda\omega_{0}P_{1})\sin\left(\frac{\theta}{2}t\right)+
(A-\lambda\omega_{0}P_{2})\cos\left(\frac{\theta}{2}t\right)
&&\nonumber\\
 +\mathcal{O}(\theta^{2},h^2,\lambda^2)=0 &&
 \end{eqnarray}
 with
 \begin{equation}
 B=B_{1}+B_{2},\qquad A=A_{1}+A_{2},
 \end{equation}
and the modification of Eqs. (\ref{AABB}) - (\ref{B2}) to read

\begin{eqnarray}
\label{AABB1}
B_{1}&=&(-\dot{c}\theta+\omega_{0}^{2}d-\frac{\omega_{0}^{2}h}{2}d-\frac{d\theta^{2}}{4})\zeta\\
A_{1}&=&(\dot{d}\theta+\omega_{0}^{2}c+\frac{\omega_{0}^{2}h}{2}c-\frac{c\theta^{2}}{4})\zeta\\
B_{2}&=&2\dot{d}\dot{\zeta}+d\ddot{\zeta}-c\dot{\zeta}\theta-\frac{9}{4}H^{2}d \zeta-\frac{3}{2}\dot{H}d\zeta\\
A_{2}&=&2\dot{c}\dot{\zeta}+c\ddot{\zeta}+d\dot{\zeta}\theta-\frac{9}{4}H^{2}c \zeta-\frac{3}{2}\dot{H}c\zeta~.
\end{eqnarray}
Following the same methodology  of the previous section, namely
demanding resonant behavior, one can set
\begin{eqnarray}
B_{1}&=&\lambda\omega_{0}P_{1}\\
A_{1}&=&\lambda\omega_{0}P_{2}\\
B_{2}&=&0\\
A_{2}&=&0~.
\end{eqnarray}
From the first two equations
and up to order $\mathcal{O}(\epsilon,h,\lambda)$ it follows that
\begin{eqnarray}
2sc+d\Big(\frac{h\omega_{0}}{2}+\epsilon\Big)=-\lambda\Big(\frac{P_{1}}{\zeta_{0}}\Big), \la{pr99} \\
2sd+c\Big(\frac{h\omega_{0}}{2}-\epsilon\Big)=\lambda\Big(\frac{P_{2}}{\zeta_{0}}\Big). \la{pr100}
\end{eqnarray}
We may now assume that $\lambda\sim\epsilon^{1+p}$ with $0\leq p< 1,$ and keep
terms up to order $\mathcal{O}(\epsilon^{2}),$ we then find
\begin{equation}
\label{sone}
2s_0s_{1}=\frac{1}{4\zeta_{0}
c_{0}}\Big(\frac{hw_0}{2}+\e\Big)[RP_{1}-P_{2}]
\end{equation}
which reduces, for $\zeta_{0}=1$ when $a=1$ to Eq.(\ref{s0}).
\par\noindent
Rewriting equations $A_{2}=0$ and $B_{2}=0$ as in the previous section
yields
\begin{equation}
\label{geq}
\frac{d}{dt}(e^{2st}\dot{\zeta})-\left(\frac{9}{4}H^{2}+\frac{3}{2}\dot{H}\right)\zeta e^{2st}=0
\end{equation}
where $s$ is defined by
\begin{equation}
s=s_{0}+\lambda s_{1}
\end{equation}
and $s_{1}$ is given by Eq. (\ref{sone}). At this point, it should be noted that $s_{1}$ is time-dependent,
so is $s.$ Equation (\ref{geq}) may be written out as
\begin{equation}
\label{ggg}
\ddot{\zeta}+(2\lambda
\dot{s}_{1}t+2s)\dot{\zeta}-\left(\frac{9}{4}H^{2}+\frac{3}{2}\dot{H}\right)\zeta=0
\end{equation}
giving at order $\mathcal{O}(\lambda^{0})$ and $\mathcal{O}(\lambda),$ respectively,
\begin{eqnarray}
\ddot{\zeta}_{0}+2s_{0}\dot{\zeta}_{0}-\left(\frac{9}{4}H^{2}+\frac{3}{2}\dot{H}\right)\zeta_{0}\!&=&\!0\\
\label{ordr0eq}
\ddot{\zeta}_{1}+2s_{0}\dot{\zeta}_{1}-\left(\frac{9}{4}H^{2}+\frac{3}{2}\dot{H}\right)\zeta_{1}\!&=&\!\!-2(\dot{s}_{1}t+s_{1})\dot{\zeta}_{0}.
\label{order1eq}
\end{eqnarray}
The solution to the first equation has been determined in the
previous section to be
\begin{equation}
\label{zerosol}
\zeta_{0}(t)=t^{\frac{3}{2}q}~.
\end{equation}
The task now is to substitute Eqs. (\ref{zerosol}) and (\ref{sone})
into Eq. (\ref{order1eq}), and ignore terms of order $\epsilon$ as
they are small compared to the others.
We employ Eq. (\ref{sone}) in order to evaluate $s_1$ but for this purpose we also need to specify the form
of functions $P_1$ and $P_2$.
%
%
%
This would determine explicitly function $\zeta_{1}$ and thus the complete resonant
solutions in an expanding background where nonlinearities are present will be of the form
\begin{equation}
\label{solution}
Y^{\pm}(t)\!\propto\! e^{[\pm
s_{0}+\frac{3q}{2t}\!\ln t + \lambda (s_{1}+\frac{\zeta_{1}}{\zeta_{0} t})]t}
[\cos\!\!\left(\!\frac{\theta}{2}t\!\right)\! \mp\! R\!\sin\!\!\left(\!\frac{\theta}{2}t\!\right)\!]
\end{equation}
where $s_{1},$ $\zeta_{0},$ and $\zeta_{1}$  are all known for a given
nonlinearity expressed by $P_1$ and $P_2$.
%
%
%
\subsection{Parametric Resonance in an expanding background with GUP}
%
%
%
\par\noindent
In this subsection we consider the nonlinearity to be the GUP term, namely,
\begin{equation}
\lambda
P=-\kappa_{1}\omega_{0}^{2}\chi\dot{\chi}^{n-2}
\end{equation}
with
\begin{equation}
\chi=a^{-3/2}Y,\qquad \dot{\chi}=a^{-3/2}(\dot{Y}-\frac{3}{2}H Y)~.
\end{equation}
In order to find the expressions of $P_{1}$ and $P_{2}$
defined in Eq. (\ref{PP}), one can follow the steps outlined in section V,
and simply make the following replacements
 \begin{eqnarray}
 &a&\rightarrow a^{-3/2} c \zeta,\quad
 b\rightarrow a^{-3/2} d \zeta,\quad
 c\rightarrow  \frac{1}{2} a^{-3/2}(c-i d) \zeta, \nonumber \\
 &A&\rightarrow B=
 a^{-3/2} \left( \bar{A}-\frac{3}{4}H(c-i d)\zeta\right)~,
 \label{B}
 \end{eqnarray}
 with
 $\bar{A}=\frac{1}{2}[(\dot{c}\zeta + c\dot{\zeta}+d\zeta\frac{\theta}{2})-i
 (\dot{d}\zeta+d\dot{\zeta}-c\zeta\frac{\theta}{2}).$
 It should be stressed that the quantity ``$a$" that appears on the right-hand side of the above replacements is the scale factor
 of the expanding universe.
 Thus the expressions of  $P_{1}$ and $P_{2}$ read
\bea
 P_1& =&\label{Pone}\! \ell_{0}\, a^{-9/2}  c_{0}^{3} \zeta_{0}^{3}\nonumber\\
 &\times&\!\!\!\!\!\!\le[ \frac{n}{n-2}(4s_0^2+\theta^{2})R+R(4s_0^2-\th^2)-4 s_0 \th
\!\ri]~~~~ \\
\label{Ptwo}
P_2 & =&\!\ell_{0}\, a^{-9/2} c_{0}^{3} \zeta_{0}^{3}\nonumber\\
 &\times&\!\!\!\!\!\!\le[ \frac{n
}{n-2}(4s_0^2+\theta^{2})+(4s_0^2-\theta^{2}) +4 Rs_0\th
\ri] \\
\mbox{with}&&\!\!\! R=\frac{d_{0}}{c_{0}}~,~~\ell_{0} = \frac{\k^{(0)}_2  (1+R^2)}{16}~,\\
\mbox{and}&&\!\!\!\!\!\! \k^{(0)}_2 = -\k n (n-1) m^{n-2}\o_0 {{n-2}\choose {n/2}}\! |B_{0}|^{n-4}
\eea
where $B_{0}$ is the quantity $B$ as defined in Eq. (\ref{B}) but in which $c, d ,\zeta$
have been replaced with $c_{0}, d_{0} ,\zeta_{0}$, respectively.
\par\noindent
Note that these two expressions reduce respectively to Eq. (\ref{fone}) and Eq. (\ref{ftwo}) when
we take the limit $H=0,\,a(t)=1,\, \mbox{and}\,\, \zeta_{0}(a)=1$.
Furthermore, the expression  of $s_{1}$ as given in Eq. (\ref{s0}) now takes the form
\begin{eqnarray}
\label{S1}
s_1 &=& \frac{\ell_{0} a^{-9/2} c_{0}^{2} \zeta_{0}^{2}}{4s_0 }\Big(\frac{h\o_0}{2}+\e\Big)
 \Big[(R^2 - 1)\Big(\frac{n}{n-2}(4
s_0^2+\th^2)\nonumber\\&&+(4s_0^2-\th^2)\Big)-8 R s_0\th\Big]
\end{eqnarray}
where $n$ can be equal to $4$ or $3$ depending on the version of GUP under consideration.
In Section V was shown that when n is even one obtains the dominant
resonance terms, while when n is odd one gets the higher order resonance terms which are weaker.
Thus, employing $c_{0}=C_{0}e^{s_{0}t}$, $a(t)=t^{q}$, $H=qt^{-1}$, $\zeta_{0}=t^{\frac{3q}{2}},$ and
keeping only terms of order $\mathcal{O}(\e^{0})$, Eq. (\ref{S1}) for $n=4$ becomes
\begin{equation}
\label{sonef}
s_{1}= -\frac{3\kappa m^2 \omega_{0}^{3}}{4s_0}\left(R^{4} - 1 \right)\left(\frac{h\omega_{0}}{2}+\epsilon \right)
{C_{0}}^{2}\,e^{2\,t\,{s_0}} \,t^{-\frac{3\,q}{2}}~.
\end{equation}
\par\noindent
Accordingly, Eq. (\ref{order1eq}) is now written as
\begin{eqnarray}
\label{eqf1f}
\ddot{\zeta}_{1}-\beta(q)t^{-2}\zeta_{1} &\simeq&
\frac{9\kappa m^2 \omega_{0}^{3}}{4 s_0} q \left(R^{4} - 1 \right)\left(\frac{h\omega_{0}}{2}+\epsilon \right)\nonumber\\
&\times& {C_{0}}^{2}\,e^{2\,t\,{s_0}} \,t^{-1}~.
\end{eqnarray}
%
%
%
%
%
%
%
%
%
%
%
%
%
%
%
%
The solution of Eq. (\ref{eqf1f}) for $\zeta_{1}$ together with the
expression (\ref{sonef}) of $s_{1}$ will determine the form of the
solution (\ref{solution}).
\par\noindent
Once again, from Eq. (\ref{sonef}) one can infer about the sign of $s_1$, which as in the case of a static background, depends only on the GUP considered (sign of $\kappa$) and the initial conditions on the matter field (via $R$).
\begin{center}
  \begin{tabular}{ | c || c | c| }
    \hline
     & sign of $s_1$ & sign of $s_1$  \\
     & ($\kappa>0 $) & ($ \kappa < 0 $) \\
     \hline \hline
    $R>1$  &  $-$ & $+$ \\
\hline
    $R<1$ &
$ + $ & $-$ \\
\hline
%
%
  \end{tabular}
\end{center}
It is evident that $s_1=0$ for $R=1$.
%
%
%
\subsection{Instability region}
%
%
\par\noindent
In this case, the instability region is obtained by setting the exponent $s$ of the complete resonant
solutions in an expanding background with nonlinearities (Eq. (\ref{solution})) to zero. The exponent is
%
%
\be
s = \pm
s_{0}+\frac{3q}{2t}\!\ln t + \lambda (s_{1}+\frac{\zeta_{1}}{\zeta_{0} t})~
\ee
and setting $s=0$ and using Eq. (\ref{unperturbed-s0}), one gets
\begin{eqnarray}
\epsilon_{min}&=&\!\!-\sqrt{\!\!\left(\frac{h\omega_0}{2}\right)^2 \!\! - 4\!\!\left(\frac{3q}{2 t}\ln t  + \lambda (s_{1}+\frac{\zeta_{1}}{\zeta_{0} t})\label{all1} \right)^2} \la{instA} \\
\epsilon_{max}&=&\!\!+\sqrt{\!\!\left(\frac{h\omega_0}{2}\right)^2 \!\!- 4\!\! \left(\frac{3q}{2 t}\ln t  + \lambda (s_{1}+\frac{\zeta_{1}}{\zeta_{0} t}) \label{all2}\right)^2} \la{instB}
\end{eqnarray}
with the instability region given as usual by
$\Delta \epsilon = \epsilon_{max}-\epsilon_{min}$.
A number of comments are in order here. First, it is easily seen that switching off GUP
by setting $\lambda=0$ reduces Eqs. (\ref{instA}) and (\ref{instB}) to the ones relevant for an expanding
universe without non-linearities, cf. Eqs. (\ref{backg1}) and (\ref{backg2}). Similarly, setting
$\lambda = 0$ {\it and} $q=0$ reduces the above to
ordinary PR, and the corresponding range Eq. (\ref{width1}).
Finally, although by setting $q=0$ one should in principle recover the width given by Eq. (\ref{gup}),
(by finding the solution for $\zeta_{1}$ in Eq. (\ref{eqf1f}) and using the expression for $s_1$ in
Eq. (\ref{sonef})), this is seen much more easily by setting $s=0$ in Eqs. (\ref{pr99}) and (\ref{pr100}),
leading to the instability band
%
%
%
%
%
%
%
%
%
\begin{eqnarray}
\Delta \e&=&h \o_0 - \frac{\l}{c_0 R \zeta_{0}} \le(R P_2 - P_1\ri)
\end{eqnarray}
which on using Eqs. (\ref{Pone}) and (\ref{Ptwo}) for $P_1$ and $P_2$, respectively, reduces to
\begin{eqnarray}
\label{width2}
\Delta \e &=& h \o_0  - \frac{\l  \ell_{0}\, a^{-9/2} c_{0}^{2} \zeta_{0}^{2}}{R} \Big[4 (1+R^{2})s_0 \th \Big]\nonumber\\
 &=& h \o_0  - 2 \l h \o_{0} \th \ell_{0} c_{0}^{2} \zeta_{0}^{2} a^{-9/2}     \nonumber\\
&=& h \o_0 \Big[ 1  - 2 \l \th \ell_{0} c_{0}^{2} \zeta_{0}^{2} a^{-9/2} \Big]~.
\end{eqnarray}
%
%
%
In this case too, the increase in the instability region depends on the GUP parameter, as well
as parameters related to the expansion of the background, which may further magnify the GUP effect.
Still the effect may remain small and unobservable due to powers of inverse Planck mass (via $\lambda$), although the situation can change with better experiments and observations in the future.
It is evident that setting $a(t)=1$ and $\zeta_0 =1$, and also
replacing $c_0 \rightarrow a_0$, reduces the above to Eq. (\ref{gup}).
%
%
\section{Concluding Remarks}
%
%
%
\par\noindent
In this paper, we studied the effect of GUP on parametric resonance in post-inflation preheating
in a static as well as an expanding background,
and showed that depending on the exact form of the GUP and initial conditions, the phenomenon of
parametric resonance and the corresponding instability band can increase, potentially resulting
in higher rates of particle production and an early termination of the above.
We believe the inclusion of GUP takes into account (at least partially) remnant Planck scale effects
on the reheating of the universe.
%
%
%
It would be interesting to study the effects of back-reaction of the produced particles in our set-up.
%
%
It would also be interesting to apply our approach to string-inspired models, for which nonlinearities enter the matter field equation via the Born-Infield  term, \emph{viz}, $\mathcal{L}\sim 1-\sqrt{1-\dot{\chi}^2}$. This would
perhaps shed some light on the inflationary reheating theory in the context of extended objects, e.g, D-branes.
Last, since our matter field Hamiltonian (Eq.(\ref{ham1})) is non-relativistic,
it might be worthwhile studying its relativistic generalization
(e.g. via a GUP modified Klein-Gordon equation), to have a better understanding of the massless limit among other things.
Note that our general formalism, including Eqs.(\ref{eom2}) and
(\ref{Phu}) remain well suited for such generalizations, as well as for a large class on non-linear corrections to parametric resonance and particle production.
%
We hope to report on these elsewhere.
\\
\\
\no {\bf Acknowledgments}
%
%
\par\noindent
A. Farag Ali would like to thank Francis-Yan Cyr-Racine for correspondence.
We thank the anonymous referee for useful comments which helped improve the paper.
W. Chemissany is supported in part by the FWO - Vlaanderen, Project No. G.0651.11, and
in part by the Federal Office for Scientific, Technical and Cultural Affairs through the
``Interuniversity Attraction Poles Programme - Belgian Science Policy" P6/11-P.
A. Farag Ali and S. Das are supported in part by the Natural
Sciences and Engineering Research Council of Canada and the Perimeter Institute for Theoretical Physics.
\\
%
%
%

\end{document}